\documentclass[letterpaper,twocolumn,letter]{jpsj3}
\usepackage{txfonts}

\usepackage{xcolor}

\usepackage{changes}

\usepackage{txfonts}
\usepackage{bm}

\definechangesauthor[name = Dai, color=red]{DA}

\title{High Field Superconducting Phases of Ultra Clean Single Crystal UTe$_2$}

\author{
Dai~Aoki$^1$\thanks{E-mail: aoki@imr.tohoku.ac.jp}, 
Ilya~Sheikin$^2$,
Nils~Marquardt$^{2,3}$,
Gerard~Lapertot$^3$,
Jacques~Flouquet$^{1,3}$, and
Georg~Knebel$^3$,
}

\inst{%
$^1$IMR, Tohoku University, Oarai, Ibaraki 311-1313, Japan\\
$^2$LNCMI-EMFL, CNRS, UGA, 38042 Grenoble, France\\
$^3$Univ. Grenoble Alpes, CEA, Grenoble INP, IRIG, PHELIQS, F-38000 Grenoble, France\\
}

\abst{%
We report the magnetoresistance of high-quality single crystals of UTe$_2$ with $T_{\rm c}=2.1\,{\rm K}$ in high magnetic fields up to $36\,{\rm T}$, with the field direction between the $b$ and $c$-axes. 
From the angular dependence of the upper critical field $H_{\rm c2}$, we found that the field-reentrant superconducting phase near $H\parallel b$-axis extends up to a field angle ($\theta\sim 24\,{\rm deg}$) from the $b$ to $c$-axis,
where another field-reentrant superconducting phase begins to appear above the metamagnetic transition field, $H_{\rm m}$. 
Our results suggest that the field-reentrant superconductivity below $H_{\rm m}$ near the $b$-axis is closely related to the superconductivity above $H_{\rm m}$ when the field is tilted toward the $c$-axis.
Superconductivity appears to be robust when the field direction is maintained perpendicular to the magnetization easy axis, 
implying that fluctuations boosting superconductivity may persist.
At first glance, these findings resemble the field-reentrant (reinforced) superconductivity observed in ferromagnetic superconductors URhGe and UCoGe, where Ising-type ferromagnetic fluctuations play a crucial role.
However, in UTe$_2$, the fluctuations are more complex.
The angular dependence of the upper critical field $H_{\rm c2}$ contrasts with that of the initial slope of $H_{\rm c2}$ near $T_{\rm c}$, revealing the anisotropic field response of fluctuations.
Thanks to the high-quality samples, quantum oscillations were detected for field directions close to the $c$-axis using magnetoresistance (Shubnikov-de Haas effect) and torque (de Haas-van Alphen effect) measurements.
The angular dependence of frequencies is in good agreement with those observed previously using the field-modulation technique, confirming quasi-two dimensional Fermi surfaces.
}

\begin{document}
\maketitle

UTe$_2$ attracts much attention because of unusual superconducting properties~\cite{Ran19,Aok19_UTe2,Aok22_UTe2_review}.
UTe$_2$ has a paramagnetic ground state with a heavy electronic state associated with the Sommerfeld coefficient $\gamma \sim 120\,{\rm mJ\,K^{-2}mol^{-1}}$.
Superconductivity occurs at $T_{\rm c} \sim 1.6$--$2.1\,{\rm K}$.
It had been suggested that UTe$_2$ would be an end member of ferromagnetic superconductors, such as UGe$_2$~\cite{Sax00}, URhGe~\cite{Aok01} and UCoGe~\cite{Huy07}.
However, no ferromagnetic fluctuations are experimentally established in UTe$_2$. 
Alternatively, antiferromagnetic fluctuations with an incommensurate wave-vector were detected in inelastic neutron scattering experiments.~\cite{Dua20,Kna21}.
Furthermore, an antiferromagnetic order appears above the critical pressure, $P_{\rm c}\sim 1.5\,{\rm GPa}$~\cite{Kna24,Li21}. 
Thus, the magnetic properties and the mechanism of superconductivity for UTe$_2$ are more complicated, compared to those for ferromagnetic superconductors.

The highlight of UTe$_2$ is that superconductivity is quite robust against magnetic field, and the superconducting upper critical field, $H_{\rm c2}$, highly exceeds the Pauli limit in all field directions.
Moreover, when the field is applied along the hard-magnetization $b$-axis, field-reentrant superconductivity is realized,
and it is abruptly suppressed due to the first-order metamagnetic transition at $H_{\rm m}\sim 35\,{\rm T}$~\cite{Kne19,Ran19_HighField}.
This huge $H_{\rm c2}$ associated with the field-reentrant behavior is consistent with a scenario for spin-triplet superconductivity.
Another strong support for the spin-triplet state is the emergence of multiple superconducting phases under pressure~\cite{Bra19,Aok20_UTe2} and at high fields~\cite{Rosuel23}, which might be a consequence of spin and orbital degrees of freedom in spin-triplet superconductivity.

The angular dependence of $H_{\rm c2}$ is also remarkable.
By rotating the field direction from $b$ to $c$-axis, maintaining the field perpendicular to the easy-magnetization $a$-axis,
$H_{\rm m}$ increases with the field angle $\theta$, following the $1/\cos\theta$ dependence.
Re-entrant superconductivity disappears at $\theta \sim 10\,{\rm deg}$, but reappears above $H_{\rm m}$ at $\theta \sim 24$--$45\,{\rm deg}$~\cite{Ran19_HighField,Hel24}.
These results were obtained using UTe$_2$ single crystals with $T_{\rm c}\sim 1.6\,{\rm K}$. 

Recently, the angular dependence of $H_{\rm c2}$ and $H_{\rm m}$ using a high quality single crystal with $T_{\rm c}=2.1\,{\rm K}$ at low temperatures down to $\sim 0.4\,{\rm K}$ was reported~\cite{Wu24}.
The field-reentrant superconducting phase near $H \parallel b$-axis is expanded down to $\theta\sim 21\,{\rm deg}$.

It is important to study the angular dependence of $H_{\rm c2}$ and $H_{\rm m}$ more precisely at lower temperatures using a high quality sample with $T_{\rm c}=2.1\,{\rm K}$.
In this study, we performed  magnetoresistance measurements in high quality samples by rotating the field angle, $\theta$, from $b$ to $c$-axis.
We found that the field-reentrant superconducting phase expands up to $\theta \sim 24\,{\rm deg} \sim \arctan (b/c)$, where $b$ and $c$ are lattice constants, corresponding to the $[011]$ direction in the reciprocal space,
which we describe $[011]^\ast$.
Further increasing $\theta$, the field-reentrant superconductivity is extended above $H_{\rm m}$.
Our results indicate that the field-reentrant superconductivity below $H_{\rm m}$ is highly related to that above $H_{\rm m}$ for $\theta \gtrsim 24\,{\rm deg}$.
The angular dependence of $H_{\rm c2}$ is compared to that of the initial slope of $H_{\rm c2}$ near $T_{\rm c}$,
revealing a clear contrast between them. 
Furthermore the quantum oscillations are detected near $H\parallel c$-axis as Shubnikov-de Haas (SdH) and de Haas-van Alphen (dHvA) effects, through magnetoresistance and torque measurements, respectively.
The angular dependence of the frequencies is in good agreement with those obtained by the field modulation technique~\cite{Aok22_UTe2_dHvA,Aok23_UTe2_dHvA}.

High quality single crystals were grown using the molten salt flux (MSF) and the molten salt flux liquid transport (MSFLT) methods as described in Refs.~\citen{Sak22,Aok24_UTe2_CrystalGrowth}.
Many as-grown crystals show the (001) as well as (011) plane, which can be easily cleaved,
indicating that the $[011]^\ast$ direction is a unique direction.
Sharp Laue spots in the X-ray diffraction and a sharp specific heat jump at $T_{\rm c}=2.1\,{\rm K}$ with the small residual $\gamma$-value ($\gamma_0/\gamma_{\rm N}\sim 0.03$) indicate the high quality of our samples.
The residual resistivity is $\rho_0 \sim 0.63\,\mu\Omega\!\cdot\!{\rm cm}$ and the residual resistivity ratio  is ${\rm RRR} \sim 500$, indicating also the exceptional quality.
The magnetoresistance was measured using the four-probe AC method at high fields up to $36\,{\rm T}$ using a resistive magnet and at temperatures down to $55\,{\rm mK}$ in a top-loading dilution fridge equipped with a swedish rotator.
The low-field magnetoresistance was supplementally measured at fields up to 13 and $15\,{\rm T}$ in dilution fridges.
The electrical current was applied along the $a$-axis and the samples were rotated from $H\parallel b$ to $c$-axis and from $H\parallel b$ to $a$-axis.
The torque measurements for the dHvA effect were performed close to $H\parallel c$-axis, using a cantilever at high fields up to $36\,{\rm T}$ and at low temperatures down to $0.06\,{\rm K}$.
For the initial slope of $H_{\rm c2}$, AC calorimetry measurements were performed near $T_{\rm c}$ using a AuFe-Au thermocouple for the field directions between $a$, $b$ and $c$-axes with a rotator.

\begin{figure}[bth]
\begin{center}
\includegraphics[width= 0.9\hsize,clip]{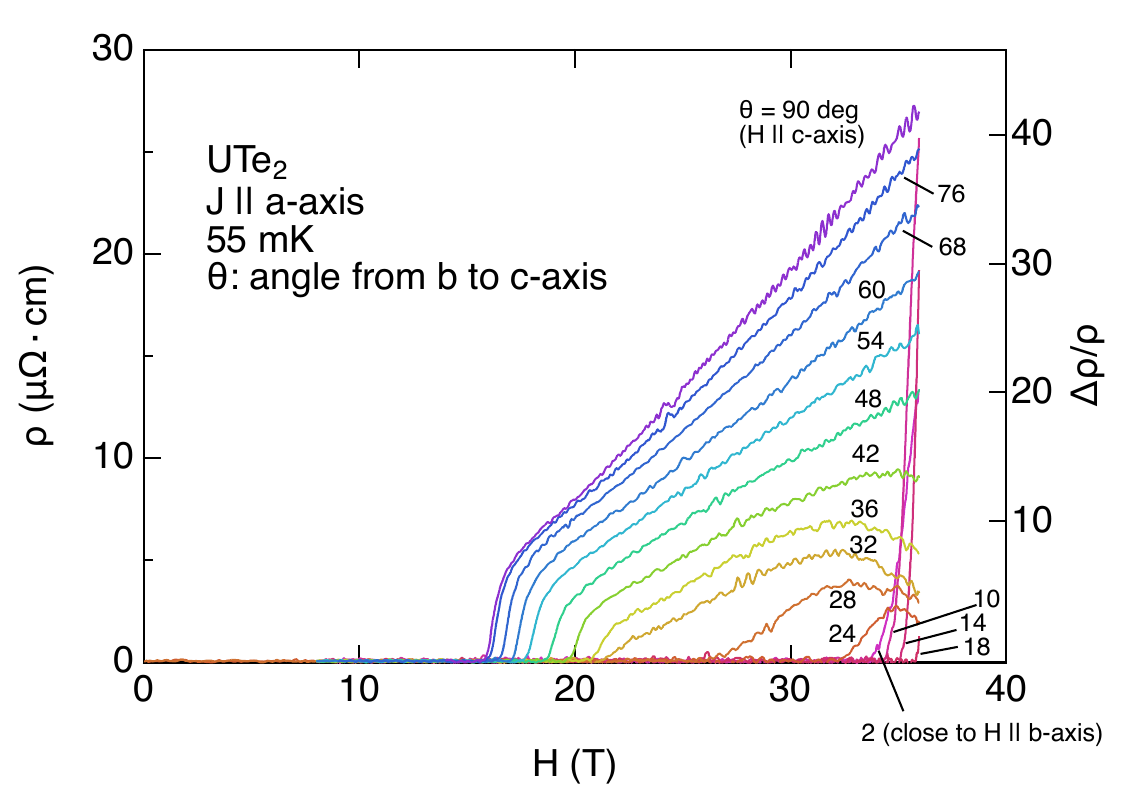}
\end{center}
\caption{(Color online) Magnetoresistance at $55\,{\rm mK}$ at different field angles tilted from $H \parallel b$ to $c$-axis in UTe$_2$. The electrical current is applied along $a$-axis.
The right axis indicates $\Delta \rho/\rho \equiv [\rho(H)-\rho(0)]/\rho(0)$, where $\rho(0)$ is the resistivity at zero field expected in the normal state.}
\label{fig:fig_MR}
\end{figure}
Figure~\ref{fig:fig_MR} shows the magnetoresistance at $55\,{\rm mK}$ at different field directions from $b$ to $c$-axis with the current along $a$-axis, maintaining the transverse configuration.
For the field direction close to $b$-axis ($\theta \sim 2\,{\rm deg}$),
superconductivity survives up to $H_{\rm c2} = 34\,{\rm T}$ and the resistivity starts to increase because of the recovery to the normal state associated with the metamagnetic transition.
Note that the normal state is not fully recovered at our highest field, $36\,{\rm T}$.
Increasing $\theta$ slightly, the onset field of magnetoresistance also slightly increases due to the increase of $H_{\rm m}$, which follows the $1/\cos\theta$-dependence.
But at $\theta=24\,{\rm deg}$, the magnetoresistance shows the finite value already at $32\,{\rm T}$ and then decreases again above $34\,{\rm T}$,
indicating the sign of reappearance of superconductivity at high fields.
The decrease of magnetoresistance at high fields is observed up to $\theta=42\,{\rm deg}$,
indicating the field-reentrant superconductivity is likely to appear at high fields above $36\,{\rm T}$ at least for $\theta \sim  24$--$42\,{\rm deg.}$
Further increasing $\theta$, $H_{\rm c2}$ decreases and the resistivity in the normal state at $36\,{\rm T}$ increases.
For $H \parallel c$-axis, the value of $H_{\rm c2}$ defined by zero resistivity is $15.9\,{\rm T}$.
Note that SdH oscillations are already visible without subtracting the background for $H\parallel c$-axis above $20\,{\rm T}$, and the magnetoresistance is large, ($\Delta \rho/\rho > 40$), confirming the high quality of our sample.
The large (small) magnetoresistance along $c$ ($b$)-axis for the current along $a$-axis 
indicates the existence of open orbits along $c$-axis in this compensated metal,
which is consistent with the cylindrical Fermi surfaces along $c$-axis detected by previous dHvA experiments.~\cite{Aok22_UTe2_dHvA,Aok23_UTe2_dHvA,Eat23}

\begin{fullfigure}[tbh]
\begin{center}
\includegraphics[width= 0.7\hsize,clip]{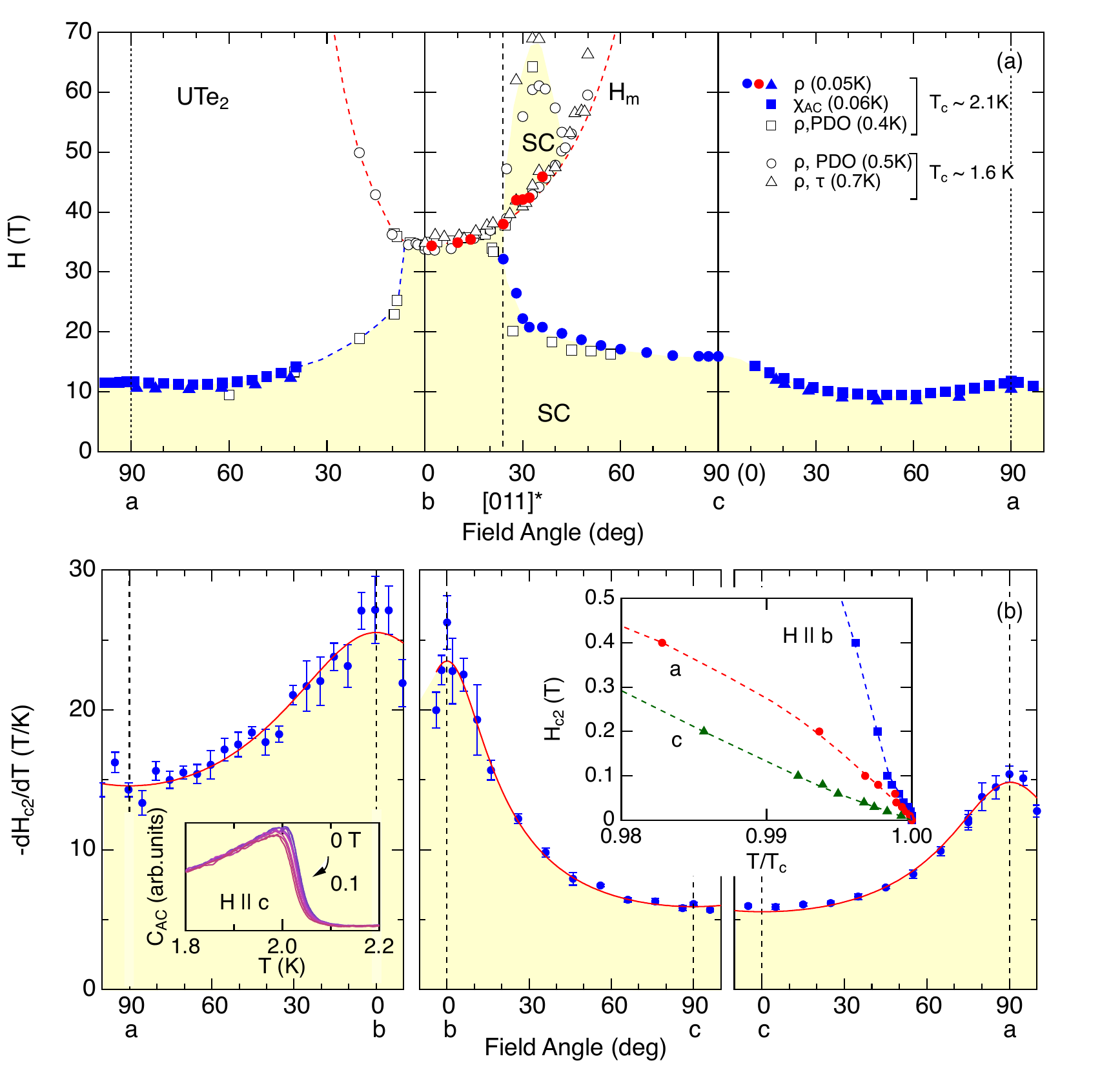}
\end{center}
\caption{(Color online) (a) Angular dependence of $H_{\rm c2}$ and the metamagnetic transition field $H_{\rm m}$ in UTe$_2$ with high quality ($T_{\rm c}\sim 2.1\,{\rm K}$). 
Closed circles and triangles are the results of magnetoresistance using a resistive and superconducting magnet, respectively.
Closed squares are the results of AC susceptibility~\cite{Aok22_UTe2_dHvA}.
Open squares are the results of magnetoresistance and TDO at $\sim 0.4\,{\rm K}$~\cite{Wu24}.
High field results above $35\,{\rm T}$ in lower quality samples ($T_{\rm c}\sim 1.6\,{\rm K}$) are also plotted.
Open circles are the results of magnetoresistance and TDO at $\sim 0.5\,{\rm K}$~\cite{Ran19_HighField}.
Open triangles are the results of magnetoresistance and torque at $\sim 0.7\,{\rm K}$~\cite{Hel24}.
(b) Angular dependence of the initial slope of $H_{\rm c2}$ near $T_{\rm c}$ determined by the AC calorimetry measurements in the field range between $0$ to $0.1\,{\rm T}$ in UTe$_2$. The error bars are from the linear fitting.
Red lines are the results of the effective mass model.
The right-top inset shows the $H_{\rm c2}$ curves near $T_{\rm c}$ for $H \parallel a$, $b$ and $c$-axis.
Temperature is scaled by $T_{\rm c}$.
The left-bottom inset shows the AC calorimetry at 0, 0.01, 0.02, 0.03, 0.04, 0.06, 0.08 and $0.1\,{\rm T}$ for $H \parallel c$-axis.
}
\label{fig:fig_Hc2_Ang3}
\end{fullfigure}
Figure~\ref{fig:fig_Hc2_Ang3}(a) shows the complete angular dependence of $H_{\rm c2}$ and the metamagnetic transition field $H_{\rm m}$.
The present results obtained by the magnetoresistance (closed circles and triangles) are combined with our results of AC susceptibility (closed squares) reported previously~\cite{Aok22_UTe2_dHvA} and the results of magnetoresistance, PDO and torque measurements around $0.4\,{\rm K}$ (open symbols)~\cite{Hel24,Ran19_HighField,Wu24}.
For the field direction close to $b$-axis, $H_{\rm c2}$ is limited by $H_{\rm m}$, which increases with the field angle, $\theta$, following the $1/\cos\theta$-dependence.
At $\theta\sim 24\,{\rm deg}$, $H_{\rm c2}$ is remarkably dropped, while $H_{\rm m}$ continuously increases.
Interestingly,  field-reentrant superconductivity starts to appear above $\sim 24\,{\rm deg}$, and persists up to $\sim 50\,{\rm deg}$~\cite{Hel24,Ran19_HighField,Wu24}
As shown in Fig.~\ref{fig:fig_MR}, the sign of field-reentrant superconductivity above $H_{\rm m}$ is observed at high fields, and the linear extrapolation of the decrease of magnetoresistance gives the starting field of reentrant superconductivity,
 which agrees well with the previously reported values~\cite{Ran19_HighField,Hel24}.

The angular dependence shown in Fig.~\ref{fig:fig_Hc2_Ang3}(a) differs somewhat from that observed in the lower quality sample ($T_{\rm c}\sim 1.6\,{\rm K}$).
In the samples with $T_{\rm c}\sim 1.6\,{\rm K}$, field-reentrant superconductivity near $b$-axis is
terminated at $\theta \sim 10\,{\rm deg}$~\cite{Kne19},
and field reentrant superconductivity above $H_{\rm m}$ for $\theta \sim 24$--$40\,{\rm deg}$ is isolated.
However, in high quality samples with $T_{\rm c}\sim 2.1\,{\rm K}$, both field-reentrant superconductivities are almost connected at $\theta \sim 24\,{\rm deg}$ at $H_{\rm m}\sim 40\,{\rm T}$.
Here we choose a rather strict definition for superconductivity, which is given by the zero resistivity.
Field-reentrant superconductivity below $H_{\rm m}$ may further extend up to $\sim 40\,{\rm deg}$ if we take a loose definition.

The angular dependences of $H_{\rm c2}$ for $H\parallel b \to a$-axis and $H\parallel c \to a$-axis are also unusual.
For instance, $H_{\rm c2}$ between $c$ and $a$-axes shows a broad minimum at $\theta \sim 45$--$60\,{\rm deg}$
with a sharp maximum at $H\parallel a$-axis.
Of course, this cannot be explained by a conventional effective mass model,
implying the anisotropic field-response of fluctuations or a multicomponent of order parameters~\cite{Zhi90}.

\begin{figure}[htb]
\begin{center}
\includegraphics[width= 0.8\hsize,clip]{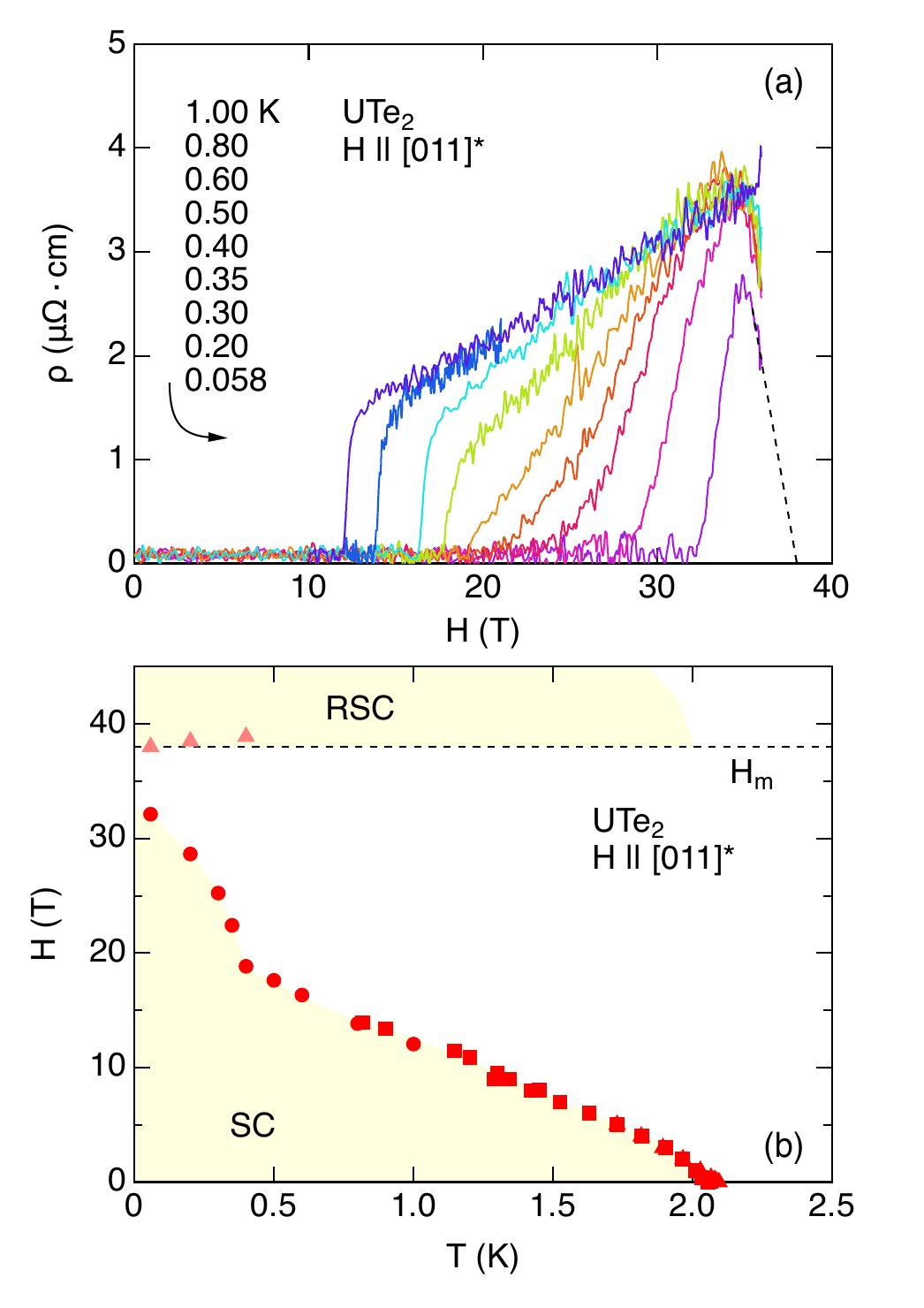}
\end{center}
\caption{(Color online) (a) Magnetoresistance at different temperatures for the field along the $[011]^\ast$ direction (field angle: $\theta\sim 24\,{\rm deg}$ from $b$ to $c$-axis) in UTe$_2$. (b) $H$--$T$ phase diagram for the field along the $[011]^\ast$ direction. Closed circles and closed squares are the results of magnetoresistance defined by zero resistivity in the resistive and superconducting magnet, respectively. Closed triangles are the estimated critical fields for reentrant superconductivity from the linear extrapolation of magnetoresistance,
which is shown for instance, as a dashed line at $0.058\,{\rm K}$ in panel (a).}
\label{fig:fig_MR_011}
\end{figure}
In order to clarify the $H$-$T$ phase diagram at the border of two reentrant superconducting phases below/above $H_{\rm m}$,
we measured the magnetoresistance at $\theta\sim 24\,{\rm deg}$ ($H \parallel [011]^\ast$) at different temperatures, as shown in Fig.~\ref{fig:fig_MR_011}(a).
At the lowest temperature, $0.058\,{\rm K}$, magnetoresistance shows a finite values above $32\,{\rm T}$, 
but it deceases again above $35\,{\rm T}$, revealing the field-reentrant superconducting behavior.
Extrapolating the decrease of magnetoresistance linearly, the magnetoresistance is expected to be zero at about $38\,{\rm T}$,
which is close to the metamagnetic transition field, $H_{\rm m}\sim 38\,{\rm T}$ for $\theta\sim 24\,{\rm deg}$.
The decrease of magnetoresistance at high fields is observed up to $0.6\,{\rm K}$, suggesting reentrant superconductivity survives at least up to $0.6\,{\rm K}$.

Figure~\ref{fig:fig_MR_011}(b) shows the $H$-$T$ phase diagram for $H\parallel [011]^\ast$.
The critical field of superconductivity here is defined by zero resistivity.
The $H_{\rm c2}$ curve shows the slight convex curvature at low fields below $18\,{\rm T}$,
but the slope abruptly increases above $18\,{\rm T}$.
Finally superconductivity is suppressed at $32\,{\rm T}$. 
Further increasing fields, superconductivity reappears above $\sim 38\,{\rm T}$, which almost coincides with metamagnetic transition field, $H_{\rm m}$.
The reentrant superconducting phase above $H_{\rm m}$ will be extended to the higher temperature at $H\gtrsim H_{\rm m}$, as shown in Fig.~\ref{fig:fig_MR_011}(a), which is expected from the results obtained previously in lower quality samples.
Here we take a strict definition for superconductivity as zero resistivity.
If we define the superconducting phase by the onset of resistivity drop, 
superconducting phases below and above $H_{\rm m}$ will be merged.

Note that the $[011]^\ast$ direction could be a specific direction for superconductivity. 
Although we do not yet have a clear explanation, it is worth mentioning that the $[011]^\ast$ direction corresponds to a cleaving plane where charge density waves (CDW) and pair density waves (PDW) associated with superconductivity have been detected on the surface through STM experiments.~\cite{Gu23,Ais23}
Furthermore, it has been proposed that the $[011]^\ast$ direction may switch to the $c$-axis direction in the tetragonal phase under pressures above $4\,{\rm GPa}$ after the structural phase transition.~\cite{Hon23}.

It is interesting to compare the angular dependence of $H_{\rm c2}$ with that of the initial slope of $H_{\rm c2}$ near $T_{\rm c}$.
Figure~\ref{fig:fig_Hc2_Ang3}(b) shows the angular dependence of the initial slope of $H_{\rm c2}$ near $T_{\rm c}$,
determined by the AC calorimetry measurements, which were done at different constant fields at different field angles.
The left-bottom inset of Fig.~\ref{fig:fig_Hc2_Ang3}(b) shows an example for $H\parallel c$-axis at low fields up to $0.1\,{\rm T}$.
The initial slopes of $H_{\rm c2}$, that is, $-dH_{\rm c2}/dT$, were obtained by the linear fitting of $H_{\rm c2}$ up to $0.1\,{\rm T}$ at different field angles.
The initial slopes for $H\parallel a$, $b$, and $c$-axis are $15$, $\sim 25$, and $5.9\,{\rm T/K}$, respectively,
showing a strong anisotropy. 
For $H\parallel b$-axis, the initial slope is very large, and the slope changes rapidly with field, as shown in the right-top inset of Fig.~\ref{fig:fig_Hc2_Ang3}.
The initial slope for $H\parallel a$-axis is also large with a convex curvature.
On the other hand, the initial slope for $H\parallel c$-axis is small, and $H_{\rm c2}$ increases linearly.
The values of initial slopes are slightly different from those reported previously~\cite{Rosuel23}, probably due to the linear range of $H_{\rm c2}$ and the definition.

The small initial slope for $H\parallel c$-axis compared to those for $H\parallel a$ and $b$-axis is consistent with the quasi-2D Fermi surfaces detected by the dHvA experiments.
More precisely, the angular dependence of the initial slope should be described by the so-called effective mass model,
which can be simply described by $H_{\rm c2}(\theta) = H_{\rm c2}(90^\circ)/(\sin^2\theta + m_c^\ast/m_a^\ast \cos^2 \theta)^{1/2}$, assuming an ellipsoidal Fermi surface with anisotropic effective mass as an average between $c$ and $a$-axis.
Here one can replace $H_{\rm c2}$ as $-dH_{\rm c2}/dT$.
The same model can be applied to the anisotropy between $c$ and $ b$-axes by replacing $m_a$ with $m_b$.
The mass anisotropy obtained by fitting with an effective mass model is $7.1$ for $m_a/m_c$ and $16$ for $m_b/m_c$.
The smaller mass anisotropy for $m_a/m_c$ is consistent with the anisotropy of magnetic susceptibilities at low temperatures, where the susceptibility for $a$-axis is the largest among the principal field directions.

\begin{figure}[tbh]
\begin{center}
\includegraphics[width= 0.8\hsize,clip]{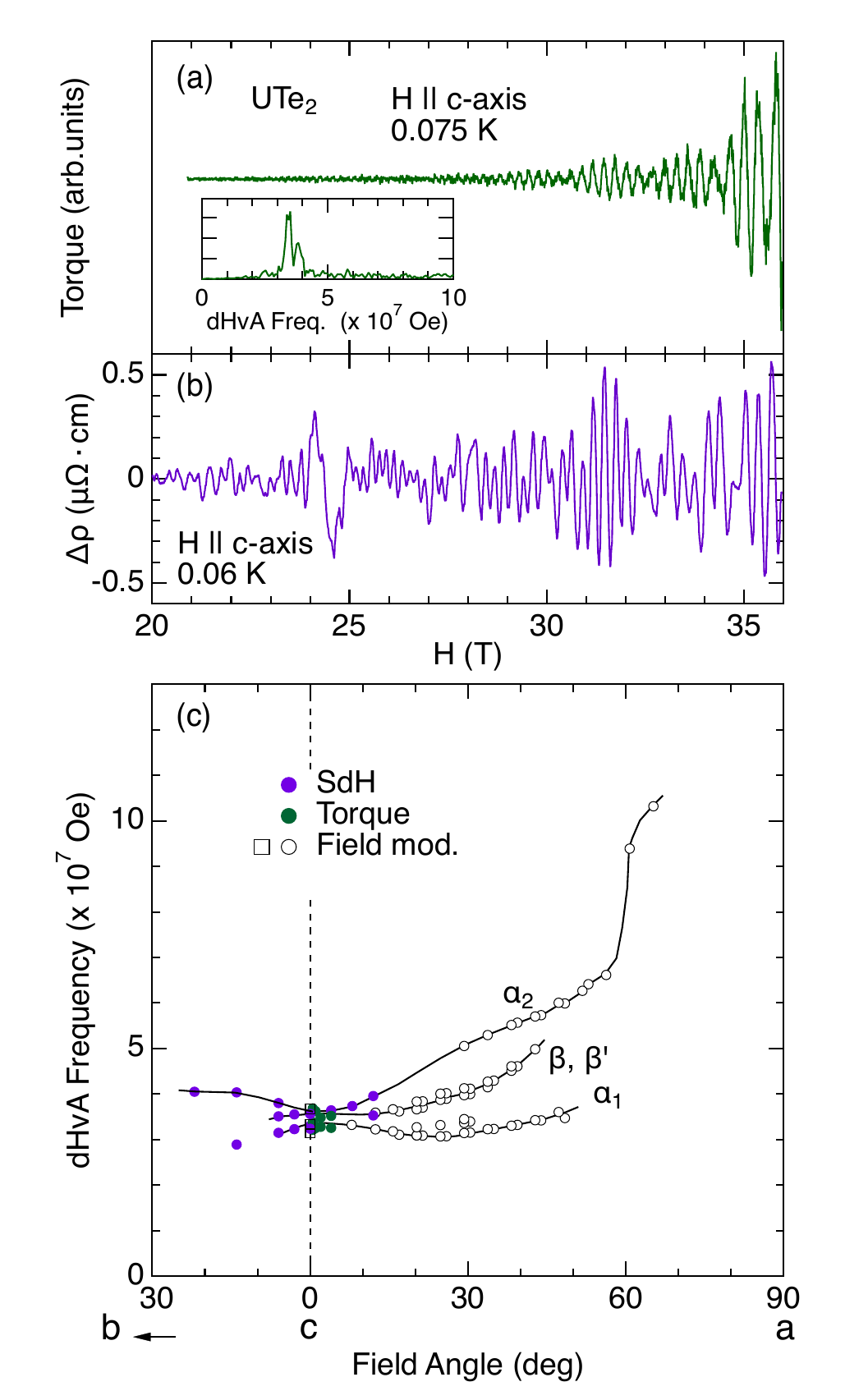}
\end{center}
\caption{(Color online) (a) dHvA oscillations and the FFT spectrum (inset) detected by torque measurements and (b) SdH oscillations for $H \parallel c$-axis in UTe$_2$ after subtracting the background signals. (c) Angular dependence of the dHvA (SdH) frequencies. Solid circles are obtained from the torque and SdH experiments. Open circles and squares are the results of our previous dHvA experiments by field-modulation technique\cite{Aok22_UTe2_dHvA,Aok23_UTe2_dHvA}.}
\label{fig:QO}
\end{figure}
Finally we show the quantum oscillations near $H\parallel c$-axis detected by magnetoresistance (SdH) and torque (dHvA) measurements.
Figures~\ref{fig:QO}(a) and \ref{fig:QO}(b) show dHvA and SdH oscillations, respectively, for $H\parallel c$-axis, after subtracting the background signals.
Thanks to high quality of the samples, clear quantum oscillations are detected above $20\,{\rm T}$ for SdH effect and above $25\,{\rm T}$ for dHvA effect.
The FFT analysis yields three fundamental frequencies ranging from $3.3$ to $3.8\,{\rm kT}$, 
which correspond to branches $\alpha_1$, $\beta$ and $\alpha_2$, respectively.
The cyclotron effective masses are $33$--$34\,m_0$, which are also in good agreement with the previously obtained values~\cite{Aok22_UTe2_dHvA,Aok23_UTe2_dHvA}.

The angular dependence of dHvA (SdH) frequencies are shown in Fig.~\ref{fig:QO}(c), together with the dHvA frequencies obtained by the previous experiments through the field-modulation technique.~\cite{Aok22_UTe2_dHvA,Aok23_UTe2_dHvA}
The present results show a good agreement with the previous results, and further extend the angular dependence near $H\parallel c$-axis for the directions from $c$ to $a$-axis and from $c$ to $b$-axis.
Note that we did not detect any low frequencies other than these fundamental frequencies in difference to Ref.~\citen{Bro23}

In summary, we performed high field magnetoresistance up to $36\,{\rm T}$ for the field direction from $b$ to $c$-axis using the high-quality single crystals of UTe$_2$ with $T_{\rm c}=2.1\,{\rm K}$.
The overall angular dependence of $H_{\rm c2}$ shows higher values corresponding to higher $T_{\rm c}$, compared to those obtained in lower-quality samples.
For the field directions close to the $b$-axis, superconductivity is cut off by the first-order metamagnetic transition at $H_{\rm m}$, insensitive to the sample quality.
Remarkably, the field-reentrant superconducting phase near $H\parallel b$-axis extends to the field angle $\theta \sim 24\,{\rm deg}$, where reentrant superconductivity starts to appear above $H_{\rm m}$.
Superconductivity is quite robust when the field direction is maintained perpendicular to the easy magnetization $a$-axis.
These properties appear to be similar to those observed in ferromagnetic superconductors, URhGe~\cite{Lev07} and UCoGe~\cite{Aok09_UCoGe}, in which ferromagnetic fluctuations are not suppressed or even induced at high fields~\cite{Tok15,Hat12}.
However, in UTe$_2$, magnetic fluctuations or other fluctuations are likely to be more complicated, in addition, associated with the first-order metamagnetic transition, which could make a drastic change of Fermi surfaces.
Another key issue is the interplay between ferromagnetic and antiferromagnetic fluctuations,
which may have quite different field-response.
The angular dependence of the initial slope of $H_{\rm c2}$ reflects the quasi-2D Fermi surfaces. 
The quantum oscillations detected by SdH and torque dHvA experiments provided several additional data points near $H \parallel c$-axis in the angular dependence of the oscillatory frequencies, which are in good agreement with the previous results measured at lower fields.

\section*{Acknowledgements}
We thank J. P. Brison, D. Braithwaite, T. Helm, M. Kimata, Y. Homma, D. X. Li, A. Nakamura, Y. Shimizu and A. Miyake
for fruitful discussion and technical support.
This work was supported by 
KAKENHI (JP19H00646, JP20K20889, JP20H00130, JP20KK0061, JP22H04933, JP24H01641), 
the French ANR projects, SCATE (ANR-22-CE30-0040), FRESCO (ANR-20-CE30-0020), FETTOM (ANR-19-CE30-0037)
We acknowledge support from
LNCMI-CNRS, member of the European Magnetic Field Laboratory (EMFL), and the Laboratoire d'excellence LANEF (ANR-10-LABX-0051).



\end{document}